\documentclass[twocolumn,aps]{revtex4-1}
\usepackage{lineno,amsmath,bm,color,array,hyperref,graphicx,url,xcolor}
\usepackage[title]{appendix}
\renewcommand{\=}{\!=\!}
\newcommand{\0}{^{\mbox{\tiny (0)}}}
\newcommand{\1}{^{\mbox{\tiny (1)}}}
\newcommand{\2}{^{\mbox{\tiny (2)}}}
\newcommand{\n}{^{\mbox{\tiny (n)}}}
\newcommand{\B}[1]{{\bm{#1}}}

\newcommand{\GR}{\ensuremath{_{\hbox{\tiny GR}}}}

\hypersetup{
   colorlinks,
   linkcolor={blue!100!black},
   citecolor={green!50!black},
   urlcolor={blue!80!black}
}

\begin{document}

\title{Non-monotonicity of the frictional bimaterial effect}
\author{Michael Aldam$^{1}$, Shiqing Xu$^{2}$, Efim A. Brener$^{3}$, Yehuda Ben-Zion$^{4}$ and Eran Bouchbinder$^{1}$}
\affiliation{$^{1}$ Chemical Physics Department, Weizmann Institute of Science, Rehovot 7610001, Israel\\
$^{2}$ National Research Institute for Earth Science and Disaster Resilience, Tsukuba, Ibaraki 305-0006, Japan\\
$^{3}$ Peter Gr\"unberg Institut, Forschungszentrum J\"ulich, D-52425 J\"ulich, Germany\\
$^{4}$ Department of Earth Sciences, University of Southern California, Los Angeles, CA 90089, USA}


%

\begin{abstract}
Sliding along frictional interfaces separating dissimilar elastic materials is qualitatively different from sliding along interfaces separating identical materials due to the existence of an elastodynamic coupling between interfacial slip and normal stress perturbations in the former case. This bimaterial coupling has important implications for the dynamics of frictional interfaces, including their stability and rupture propagation along them. We show that while this bimaterial coupling is a monotonically increasing function of the bimaterial contrast, when it is coupled to interfacial shear stress perturbations through a friction law, various physical quantities exhibit a {\em non-monotonic} dependence on the bimaterial contrast. In particular, we show that for a regularized Coulomb friction, the maximal growth rate of unstable interfacial perturbations of homogeneous sliding is a non-monotonic function of the bimaterial contrast, and provide analytic insight into the origin of this non-monotonicity. We further show that for velocity-strengthening rate-and-state friction, the maximal growth rate of unstable interfacial perturbations of homogeneous sliding is also a non-monotonic function of the bimaterial contrast. Results from simulations of dynamic rupture along a bimaterial interface with slip-weakening friction provide evidence that the theoretically predicted non-monotonicity persists in non-steady, transient frictional dynamics.
\end{abstract}
\maketitle

\section{Introduction}
\label{sec:intro}

Bimaterial configurations with two solids of different elastic properties in contact exist widely in earth and man-made systems. One class of examples involves large strike-slip faults that separate different crustal blocks, as reported for the San Andreas, Hayward and San Jacinto faults in California~\citep{Ben-Zion1991,Eberhart-Phillips1993,Allam2014,Share2017} and the North Anatolian fault in Turkey~\citep{Sandvol2001,Ozakin2012}. Historic records and recent observations indicate that such strike-slip faults are able to produce earthquakes with magnitude up to 8~\citep{Scholz2002,Rockwell2015,Bohnhoff2016}. The contrast of seismic properties across large faults may become more prominent with increasing fault displacement. In addition, large faults tend to nucleate along prominent pre-existing sutures as reported for the North Anatolian fault~\citep{Sengor2005}. In a different tectonic regime, subduction zone plate interfaces juxtapose continental crust against denser oceanic crust~\citep{Turcotte2014} and host megathrust earthquakes with magnitude that may be above 9~\citep{Lay2015}. The appearance in subduction setting of compliant sediments in the overriding plate near the toe can contribute significantly to causing large slip near the trench and anomalous tsunami excitation during large earthquakes~\citep{Satake1999,Scholz2014,Lotto2017}. Bimaterial fault configurations may also develop spontaneously from initially homogeneous solids by damage mechanisms that create a low velocity layer~\citep{Yamashita2000,Lyakhovsky2011,Xu2015,Thomas2017}. Once such layer is formed, future slip events or deformation bands tend to migrate to the boundary between the low velocity zone and relatively intact host rocks~\citep{Brietzke2006,Huang2014}. Another important class of bimaterial configurations is ice-rock interfaces as in the Whillans Ice Plain region of the Western Antarctica~\citep{Wiens2008,PaulWinberry2013,Lipovsky2016}. Bimaterial interfaces exist also at the bottom of landslides and may be relevant for the continuing propagation of long-runout landslides in areas with little to no topographic gradient~\citep{BrianDade1998,Legros2002}. Other examples include interfaces in mines (edges of mined layers) and a variety of composite engineering materials and structures.

The presence of material contrast across a fault produces a generic failure instability mechanism associated with elastodynamic coupling between interfacial slip and changes in normal stress~\citep{Weertman1980}, along with head waves that propagate along the interface and contribute to dynamic changes of stress~\citep{Ben-Zion1989,Ben-Zion1990}. These effects and several geophysical paradoxes (e.g.~the heat flow paradox and the absence of rupture branching from large faults) that arise in theoretical considerations in a homogeneous solid motivated numerical simulations of bimaterial ruptures that explore parameters producing self-sustained ruptures even without frictional weakening~\citep{Andrews1997,Ben-Zion1998}. Such simulations with a constant friction coefficient (and no other intrinsic physical length scale) do not have a continuum limit, so the numerical solutions are ill-posed~\citep{Ranjith2001}. Consequently, the results contain a mixture of physical features and numerical artifacts, but careful examination of sets of solutions with different physical and numerical parameters may still allow extracting tentative conclusions on the system behavior~\citep{Ben-Zion2001}. The results from the early numerical studies suggested that bimaterial ruptures can produce narrow self-healing slip pulses with low dynamic stress (and low associated frictional heat) that may resolve key outstanding paradoxes~\citep{Ben-Zion1998}. These and later studies highlighted several additional features of bimaterial ruptures that can have significant effects on fault mechanics and on generated ground motion, including preferred rupture directivity, generation of rock damage asymmetry across faults and tensional source term with potential for fault opening and rock pulverization~\citep{Cochard2000,Ben-Zion2005,Ampuero2008a,Xu2017}. Many additional studies attempted to clarify properties of bimaterial ruptures in the presence of various nucleation procedures and friction laws~\citep{Harris1997,Ben-Zion2002,Shi2006,Rubin2007,Ampuero2008a,Erickson2016} and off-fault material yielding in relation to the background principal stress orientation~\citep{DeDontney2011,Xu2012}.

The interesting dynamical features of bimaterial ruptures led to various observational studies involving laboratory experiments and in-situ data. The former focused on measuring rupture velocity and directivity along with dynamic changes in normal stress in various settings~\citep{Xia2005,Lykotrafitis2006,Bhat2010,Shlomai2016}. The latter focused on estimating rupture directivity (e.g.~\citep{Lengline2011,Wang2011a,Calderoni2015,Ross2016}) and rock damage asymmetry (e.g.~\citep{Lewis2005,Dor2006,Dor2008,Mitchell2011}) in relation to contrast of seismic velocities across faults. The observational results generally support the relevance of bimaterial ruptures to various aspects of fault mechanics and seismic hazard. It is therefore important to quantify how dynamic bimaterial effects change with the degree of property contrast across the interface. \citet{Ben-Zion1998} found that under a constant Coulomb friction the growth rate of the maximal slip velocity in a slip pulse is a non-monotonic function of material contrast across the interface and is maximal around the level where generalized Rayleigh waves barely exist. As mentioned, however, numerical simulations with constant Coulomb friction do not have a continuum limit, so they cannot be used to derive general conclusions.

In the present work we study the dependence of dynamic bimaterial effects on the degree of bimaterial contrast using a variety of theoretical and numerical analyses. This is done by examining the stability of steady state sliding and of unsteady rupture propagation in the context of frictional laws that have well-posed continuum descriptions. Our major result is that while the bimaterial coupling between interfacial slip and normal stress changes is a monotonically increasing function of the bimaterial contrast, when it is coupled to interfacial shear stress perturbations through a friction law, various physical quantities exhibit a non-monotonic dependence on the bimaterial contrast. We also show that this non-monotonicity is not directly related to the existence/inexistence of generalized Rayleigh waves. The generic non-monotonicity of the bimaterial effect is established in a few steps. First, section~\ref{sec:bimaterial} establishes the monotonicity of the bimaterial coupling between interfacial slip and normal stress perturbations. Section~\ref{sec:FrictionLaw} outlines the basic theoretical framework that connects slip with shear and normal stress changes along a bimaterial interface once a friction law is taken into account. Sections~\ref{sec:Coulomb}-\ref{sec:RSF} introduce different types of friction laws --- a regularized Coulomb friction and a velocity-strengthening rate-and-state friction --- and show analytically and semi-analytically that the maximal growth rate of unstable linear perturbations of steady sliding exhibits a non-monotonic dependence on the degree of bimaterial contrast. Finally, section~\ref{sec:dynamics} demonstrates the existence of non-monotonic behavior in numerical simulations of transient rupture evolution under a linear slip-weakening friction law. The results and their implications are summarizes and discussed in the concluding section~\ref{sec:summary}.

\section{Monotonicity of the bimaterial coupling between interfacial slip and normal stress perturbations}
\label{sec:bimaterial}

We start by considering two linear elastic half-spaces in mechanical contact at an interface located at $y\!=\!0$. The coordinate along the interface is $x$ and plane-strain conditions are assumed. The bimaterial coupling between interfacial slip and normal stress emerges in this physical situation from the elastodynamics of the two half-spaces when they are made of different linear elastic materials, i.e.~at least one of their linear elastic moduli or mass density is different. To see how this coupling emerges, note that each half-space satisfies the Navier-Lam\'e equation $\nabla\cdot{\B \sigma}\=\frac{\mu}{1-2\nu}\nabla\!\left(\nabla\cdot{\B u}\right)+\mu\nabla^2{\B u}\=\rho\,\ddot{\B u}$, with its own shear modulus $\mu$, Poisson's ratio $\nu$, mass density $\rho$, displacement vector field $\B u$ and Cauchy stress tensor field $\B \sigma$~\citep{Landau1986}. Hooke's law was used to relate $\B \sigma$ to $\B u$, and each superimposed dot represents a partial time derivative. As the bimaterial coupling between interfacial slip and normal stress manifests itself in the presence of spatiotemporally varying fields, we express the deviations from homogeneous fields at the interface, approached from its upper and lower sides $y\=0^\pm$, by $\B \sigma(x,t)\=\delta \B \sigma\,\exp\!\left[-i k \left(z c_s t + x\right)\right]$ and $\B u(x,t)\=\delta \B u\exp\!\left[-i k \left(z c_s t + x\right)\right]$, where $k$ is the wavenumber along the interface, $-z$ is the complex dimensionless phase-velocity and $c_s\=\sqrt{\mu/\rho}$ is the shear wave-speed in each half-space. That is, $\delta \B \sigma$ and $\delta \B u$ are the Fourier amplitude of the fields.

Each of the two problems formulated above, for the upper and lower half-spaces, admits a solution that at the interface takes the form $\delta u_i\=M_{ij}(z,k)\,\delta\sigma_{yj}$. For the upper half-space --- denoted hereafter by the superscript $\mbox{\footnotesize (1)}$ --- the matrix $\B M$ takes the form~\citep{Ranjith2001}
\begin{widetext}
\begin{equation}
\label{eq:mat_G_def}
{\B M}\1(z,k)=\frac{1}{\left|k\right| \mu\,R(z,\beta)}\left(
\begin{array}{cc}
 \alpha _s \left(\alpha _s^2-1\right) & i\,\hbox{sgn}(k) \left(\alpha _s^2-2 \alpha _d \alpha _s+1\right) \\
 -i\,\hbox{sgn}(k) \left(\alpha _s^2-2 \alpha _d \alpha _s+1\right) & \alpha _d \left(\alpha _s^2-1\right) \\
\end{array}
\right)\ .
\end{equation}
\end{widetext}
Here $\alpha_s(z)\!\equiv\!\sqrt{1-z^2}$ and $\alpha_d(z,\beta)\!\equiv\!\sqrt{1-\beta^2 z^2}$, where  $\beta\!\equiv\!\frac{c_s}{c_d}\!=\!\sqrt{\frac{1-2 \nu}{2-2\nu}}$ is the ratio of the shear wave-speed $c_s$ to the dilatational wave-speed $c_d$ and $R(z,\beta)\!\equiv\!4\alpha_d\alpha_s-\left(\alpha_s^2+1\right)^2$ is the Rayleigh function~\citep{ACHENBACH1975}. For the lower half-space --- denoted hereafter by the superscript $\mbox{\footnotesize (2)}$ --- ${\B M}\2(z,k)$ is the same as ${\B M}\1(z,k)$ except that its diagonal elements have opposite signs. Up until now, the elastodynamic solutions in the two half-spaces were considered separately. The coupling between the two solutions appears when continuity and jump conditions across the interface at $y\=0$ are introduced~\citep{Geubelle1995}. Specifically, we demand continuity of the normal component of the interfacial displacement vector field (i.e.~we assume no opening gaps), $\delta{u}_y\1-\delta{u}_y\2\=0$, and of $\sigma_{yx}$ and $\sigma_{yy}$, but allow for a discontinuity in the tangential component of the interfacial displacement vector field (interfacial slip), $\delta{u}_x\1-\delta{u}_x\2\!\equiv\!\delta\epsilon$, which later on will appear in the interfacial constitutive relation. Using these interfacial conditions, one obtains~\citep{Geubelle1995}
\begin{equation}
  \delta{\B u}\1-\delta{\B u}\2=\left(
\begin{array}{c}
 \delta \epsilon  \\
 0 \\
\end{array}
\right)=
\left({\B M}\1-{\B M}\2\right)\left(
\begin{array}{c}
 \delta \sigma_{yx}  \\
 \delta \sigma_{yy} \\
\end{array}
\right) \ .
\label{eq:Mdifference}
\end{equation}
By inverting this relation and defining ${\B G}\!\equiv\!\left({\B M}\1-{\B M}\2\right)^{-1}$, we obtain $\delta \sigma_{yx}\=G_{11}\,\delta\epsilon$ and $\delta\sigma_{yy}\=G_{21}\,\delta\epsilon$. If the two half-spaces are made of {\em identical} materials, then $\B G$ is diagonal, implying that $\delta\sigma_{yy}\=0$ independently of the interfacial slip $\delta\epsilon$. Consequently, the elastodynamic bimaterial coupling between interfacial slip perturbations $\delta\epsilon$ and normal stress perturbations $\delta\sigma_{yy}$ is entirely encapsulated in the off-diagonal element $G_{21}$, which is non-zero only in the presence of bimaterial contrast.

\begin{figure*}[ht]
        \centering
        \includegraphics[width=.99\textwidth]{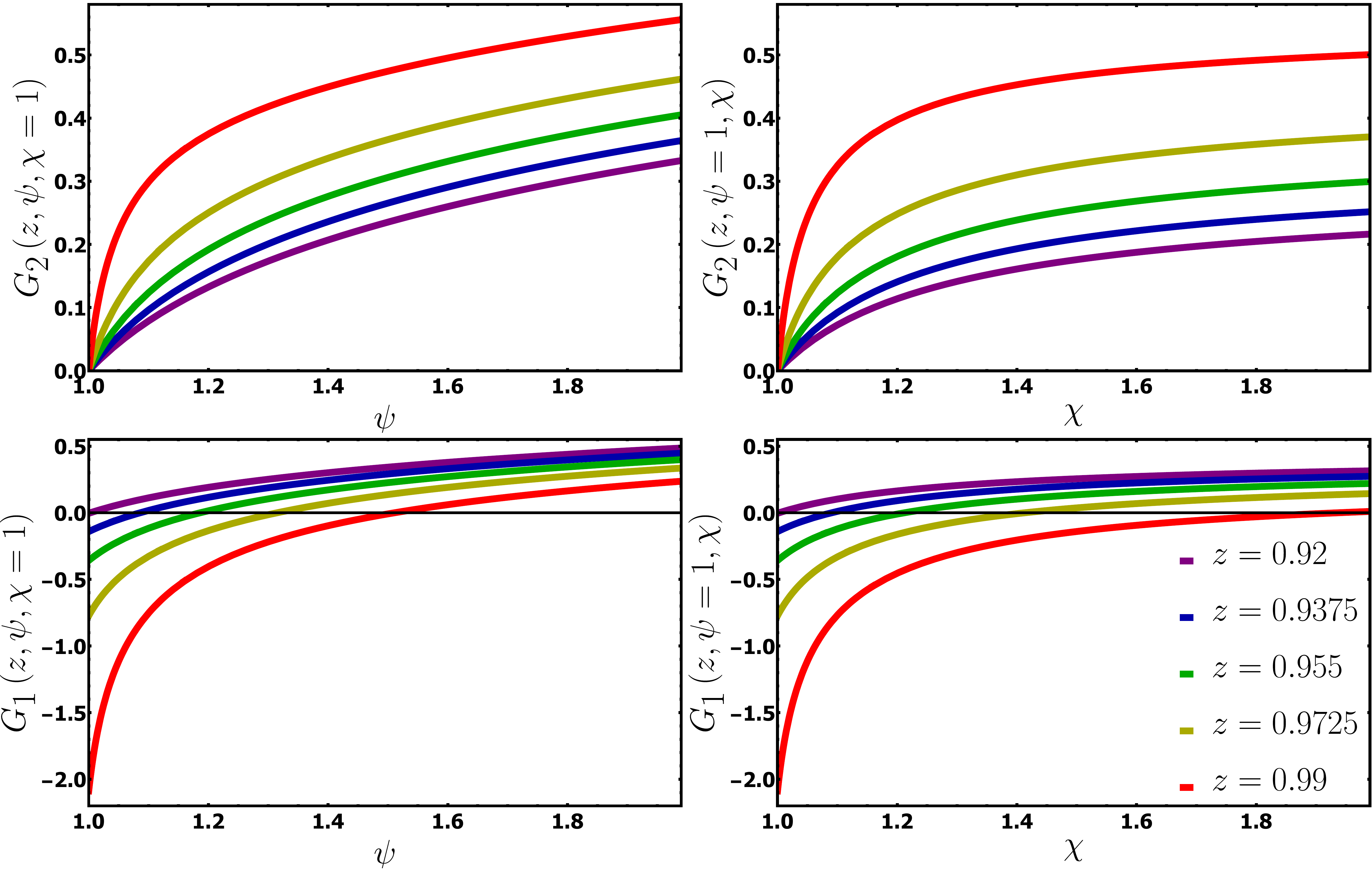}
        \caption{(Top panels) $G_2$ as a function of $\psi$ for $\chi\!=\!1$ (left) and $\chi$ for $\psi\!=\!1$ (right) for a few values of $z$ (see legend in the bottom right panel). (Bottom panels) $G_1$ as a function of $\psi$ (left) and $\chi$ (right) for the same values of $z$. In all panels $\nu\1\!=\!\nu\2\!=\!0.25$.}\label{fig:g_func}
\end{figure*}

To highlight the dependence of $G_{11}$ and $G_{21}$ on physical variables and material parameters, one needs to make a normalization choice; we choose to normalize stresses by the shear modulus of the upper half-space, $\mu\1$, and to define the complex phase-velocity $-z$ of the coupled problem with respect to the shear wave-speed of the upper half-space, $c_s\1$ (note that this normalization has already been used in both ${\B M}\1$ and ${\B M}\2$ in Eq.~\eqref{eq:Mdifference}). With these definitions, we obtain
\begin{eqnarray}
\delta\sigma_{yx}&=&\mu\1\,k\,G_1(z; \psi, \chi, \nu\1, \nu\2)\,\delta\epsilon\qquad\hbox{and}\qquad\nonumber\\
\delta\sigma_{yy}&=&-i\mu\1\,k\,G_2(z; \psi, \chi, \nu\1, \nu\2)\,\delta\epsilon \ ,
\label{eq:Gs}
\end{eqnarray}
where $G_1\!\equiv\!G_{11}/k\mu\1$, $G_2\!\equiv\!i G_{21}/k\mu\1$, $\psi\!\equiv\!\mu\2/\mu\1$ and $\chi\!\equiv\!\rho\1/\rho\2$. The explicit forms of $G_1$ and $G_2$ are presented in the Appendix. Note that $\psi$ and $\chi$ are defined such that the shear wave-speeds ratio $\zeta\!\equiv\!c_s\2/c_s\1\=\sqrt{\psi\,\chi}$ is an increasing function of both. We thus observe that the elastodynamic transfer functions $G_1$ and $G_2$, which fully characterize the elastodynamic part of the bimaterial problem, depend on the complex phase-velocity $-z$ (but not on the magnitude of the wavenumber $k$) and on $4$ dimensionless material parameters: the shear moduli ratio $\psi$, the mass densities ratio $\chi$, and the two Poisson's ratios $\nu\1$ and $\nu\2$. $G_1$ and $G_2$ are real and symmetric functions of $z$.
The first question we would like to address is the dependence of $G_2$, which quantifies the elastodynamic bimaterial coupling between interfacial slip perturbations $\delta\epsilon$ and normal stress perturbations $\delta\sigma_{yy}$, on the material contrast as quantified by $\psi$ and $\chi$. To that aim, we plot in Fig.~\ref{fig:g_func} $G_2$ as a function of $\psi$ (top left, with $\chi\=1$) and $\chi$ (top right, with $\psi\=1$), for various values of $z$ (in a range whose relevance will become clearer below) and $\nu\1\=\nu\2\=0.25$. The major observation is that $G_2$ is a monotonically increasing function of $\psi$ and $\chi$. The same behavior persists for other values of the fixed parameters and for varying $\nu\1$ when $\nu\2$ is fixed, with $\psi\=\chi\=1$ (not shown). In fact, $G_2$ is also an increasing function of $z$, as is clearly observed in Fig.~\ref{fig:g_func}. Note that, as expected, in the absence of bimaterial contrast ($\psi\!\to\!1$, $\chi\!\to\!1$ and $\nu\1\!\to\!\nu\2$), $G_2$ vanishes. We also present $G_1$ (lower panels, for the same choices of parameters), which exhibits a monotonically increasing dependence on $\psi$ and $\chi$ as well. Note that $G_1$ remains finite also in the absence of bimaterial contrast and that it is a decreasing function of $z$.

\section{Coupling interfacial normal and shear stress perturbations through a friction law}
\label{sec:FrictionLaw}

In the previous section it was demonstrated that the elastodynamic transfer functions $G_1$ and $G_2$ exhibit a monotonic dependence on the bimaterial contrast quantified by $\psi$ and $\chi$. Can we conclude then that various physical observables, especially those which identically vanish in the absence of a bimaterial contrast, generically exhibit a monotonically increasing dependence on the bimaterial contrast? We would like to argue that the answer to this question is non-trivial; that is, that the monotonic dependence of $G_1$ and $G_2$ on the bimaterial contrast parameters $\psi$ and $\chi$ does {\em not} immediately imply a monotonic dependence of any physical observable of interest on $\psi$ and $\chi$. The key point is that Eq.~\eqref{eq:Gs} is {\em not} a complete solution of the physical problem at hand because $\delta\sigma_{yx}$ and $\delta\sigma_{yy}$ are {\em not} independent; rather, they are generically coupled by an interfacial constitutive relation, i.e.~by a friction law. Moreover, the relation between the wavenumber $k$ and the complex dimensionless phase-velocity $-z$ remains unspecified in Eq.~\eqref{eq:Gs}. That is, the relations in Eq.~\eqref{eq:Gs} emerge from the bulk elastodynamics of the two half-spaces and the continuity of the normal component of the interfacial displacement vector, but still leave the relation between $\delta\sigma_{yx}$, $\delta\sigma_{yy}$ and $\delta\epsilon$ unconstrained. The relation between these quantities depends on the physics of the interface and is expressed in terms of an interfacial constitutive relation.

Quite generically, the interfacial constitutive relation --- the friction law --- can be written in the form
\begin{equation}
\dot{\tau} = F(\tau, \sigma, v, \epsilon,...) \ ,
\label{eq:FrictionLaw}
\end{equation}
where we defined $\tau(x,t)\!\equiv\!\sigma_{yx}(x,y\!=\!0,t)$, $\sigma(x,t)\!\equiv\!-\sigma_{yy}(x,y\!=\!0,t)$, $\epsilon(x,t)\!\equiv\!{u}_x\1(x,y\!=\!0,t)-{u}_x\2(x,y\!=\!0,t)$ and $v(x,t)\!\equiv\!\dot{u}_x\1(x,y\!=\!0,t)-\dot{u}_x\2(x,y\!=\!0,t)\=\dot\epsilon(x,t)$. The dots in the arguments of the function $F$ in Eq.~\eqref{eq:FrictionLaw} represent possible additional fields --- the so-called internal state fields --- which describe the structural state of the interface and satisfy their own evolution equations. Equation \eqref{eq:FrictionLaw} gives rise to coupling between the Fourier amplitudes $\delta\sigma_{yx}$, $\delta\sigma_{yy}$ and $\delta\epsilon$ appearing in Eq.~\eqref{eq:Gs}, thus fully define the physical problem at hand. The question then is whether physical quantities in this fully defined problem generically exhibit monotonic dependence on the bimaterial contrast or not. This question will be extensively addressed in what follows.

\section{Non-monotonicity of the bimaterial effect in the stability of homogeneous sliding with a regularized Coulomb friction}
\label{sec:Coulomb}

To start addressing the posed question, we first consider the following interfacial constitutive relation~\citep{Prakash1992,Prakash1993,Prakash1998}
\begin{equation}
\label{eq:PC}
\dot\tau=-\frac{v}{L}\left(\tau - f \sigma \right) \ ,
\end{equation}
which is a specific example of Eq.~\eqref{eq:FrictionLaw}. Here $f\!>\!0$ is a constant friction coefficient and $L$ is a lengthscale that controls the relaxation time $L/v$ to Coulomb friction $\tau\=f \sigma$. In the limit $L\!\to\!0$, Eq.~\eqref{eq:FrictionLaw} is reduced to Coulomb friction $\tau\=f \sigma$~\citep{Dowson1979,Persson1998}, which is known to be ill-posed (see below and~\citep{Ranjith2001}). The finite $L$ regularization renders the interfacial constitutive relation in Eq.~\eqref{eq:PC} well-posed (see below) and hence suitable for our purposes here.

To see how the interfacial constitutive relation in Eq.~\eqref{eq:PC} couples $\delta\sigma_{yx}$, $\delta\sigma_{yy}$ and $\delta\epsilon$, we consider homogeneous sliding at a velocity $V$ under the application of a compressive stress $\sigma_0$, such that $\tau_0\=f\sigma_0$. We then consider a slip displacement perturbation of amplitude $\xi$ on top of the homogeneous sliding solution, i.e.~$\epsilon(x,t)\= Vt+\xi\,\exp\left[-i k \left(t z c_s\1+x\right)\right]$, and use Eq.~\eqref{eq:Gs} to obtain
\begin{equation}
\label{eq:pert}
\begin{split}
\sigma_{yx}(x,t)&=f \sigma _0+\xi\,k\,\mu\,G_1(z)\,\exp\left[-i k \left(t z c_s\1+x\right)\right]\ ,\\
\sigma_{yy}(x,t)&=-\sigma _0-i\,\xi\,k\,\mu\,G_2(z)\,\exp\left[-i k \left(t z c_s\1+x\right)\right]\ ,
\end{split}
\end{equation}
where the parametric dependence of $G_{1,2}(z)$ on $\psi$, $\chi$, $\nu\1$ and $\nu\2$ is omitted for the ease of notation. Inserting these fields into Eq.~\eqref{eq:PC}, we obtain to leading order in $\xi$
\begin{equation}
\label{eq:spec}
\left(1-izq\right)G_1(z)-i f G_2(z)=0 \ ,
\end{equation}
where we defined the dimensionless wavenumber as $q\!\equiv\!k L c_s\1/V$. The solutions to Eq.~\eqref{eq:spec} determine the linear stability of homogeneous sliding along frictional interfaces described by the regularized Coulomb friction of Eq.~\eqref{eq:PC}. In particular, solutions with $\Im[z]\!>\!0$ are unstable, i.e.~they correspond to perturbations that exponentially grow in time, while solutions with $\Im[z]\!<\!0$ are stable, i.e.~they correspond to perturbations that exponentially decay in time.

When dealing with instabilities, one is generally interested in identifying the competing, i.e.~stabilizing and destabilizing, physical processes involved in a given problem. In our case, the relevant physics includes the regularized Coulomb friction law and the bimaterial effect. The former, which does not feature any frictional weakening behavior, is not expected to give rise to destabilization. The latter, which may give rise to a reduction in the interfacial normal stress and hence to weakening, appears to be responsible for the only potentially destabilizing process in the problem. As the bimaterial coupling between interfacial slip and normal stress variations is fully described by the elastodynamic transfer function $G_2(z)$, our strategy is to consider first the frictionless limit $f\!\to\!0$, for which the combination $f G_2$ in Eq.~\eqref{eq:spec} vanishes, and then take into account the potentially destabilizing bimaterial effect perturbatively.

For $f\!\to\!0$, Eq.~\eqref{eq:spec} admits non-trivial solutions $z\=\pm z\GR$, with a real $z\GR$, corresponding to $G_1(z\=\pm z\GR)\!=\!0$. These propagative wave solutions, known as Generalized Rayleigh waves (hence the subscript), exist for a finite range of bimaterial contrasts~\citep{Weertman1963,Achenbach1967}. For example, for $\chi\=1$, they exist in the range $0\!<\!\psi\!<\!1.85$ (for $\nu\1\=\nu\2\=0.25$) and for $\psi\=1$, they exist in the range $0\!<\!\chi\!<\!9.51$ (also for $\nu\1\=\nu\2\=0.25$). Obviously, $z\=\pm z\GR$ is {\em not} a solution of Eq.~\eqref{eq:spec}; however, when Generalized Rayleigh waves exist, we can look for solutions in the form $z\!=\!\pm z\GR+f\delta{z}$, where the complex contribution $f\delta{z}$ is treated as small~\citep{Ranjith2001}. Inserting this ansatz into Eq.~\eqref{eq:spec} and expanding to leading order in $f$, we obtain unstable solutions ($\Im[z]\!>\!0$) in the form
\begin{eqnarray}
\label{eq:lam}
z&\simeq&-z\GR+\frac{i f G_2(z\GR)}{(1+i q z\GR) \left|G_1'(z\GR)\right|}\qquad\Longrightarrow\qquad\nonumber\\
 \lambda&\equiv& q\,\Im\left[z\right] \simeq \frac{q f G_2(z\GR)}{\left(1+q^2 z\GR^2\right)\left|G_1'(z\GR)\right|}\ ,
\end{eqnarray}
where $\lambda(q)$ is the dimensionless instability growth rate.

\begin{figure*}[ht]
  \centering
  \includegraphics[width=0.99\textwidth]{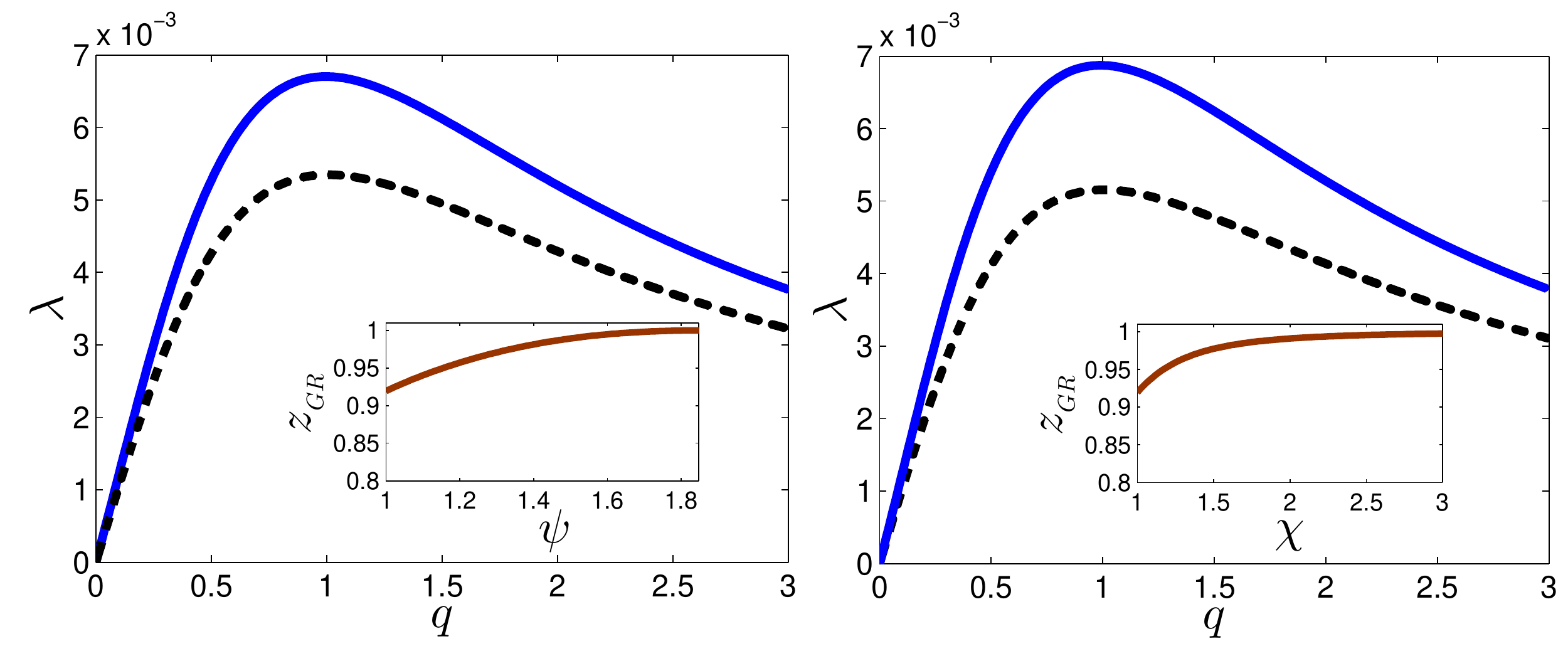}
  \caption{The instability growth rate $\lambda$ as a function of the dimensionless wavenumber $q$ (solid lines) obtained from a numerical solution of Eq.~\eqref{eq:spec} with $\psi\!=\!1.6$ and $\chi\!=\!1$ (left), and with $\psi\!=\!1$ and $\chi\!=\!2.5$ (right). The dashed lines correspond to the analytic approximation in Eq.~\eqref{eq:lam}. As the latter involves the dimensionless phase-velocity of Generalized Rayleigh waves, $z\GR$, we show in the insets the dependence of $z\GR$ on $\psi$ for $\chi\!=\!1$ (left) and on $\chi$ for $\psi\!=\!1$ (right). We used $\nu\1\!=\!\nu\2\!=\!0.25$ and $f\!=\!0.3$ to obtain the results shown in this figure.}\label{fig:PC_spec}
\end{figure*}

The analytic approximation in Eq.~\eqref{eq:lam} is compared to the full numerical solution of Eq.~\eqref{eq:spec} for two sets of parameters in Fig.~\ref{fig:PC_spec}, demonstrating favorable agreement. Note that $\lambda(q)$ features a maximum, $\lambda_{max}$, attained at $q\=q_{max}$, which corresponds to the most unstable mode. In addition, $\lambda\!\to\!0$ in the limit $q\!\to\!\infty$ (vanishingly small perturbation wavelength), which shows that indeed the problem is well-posed in the presence of a finite $L$ (ill-posedness typically corresponds to the situation in which $\lambda\!\propto\!|q|$ in the limit $q\!\to\!\infty$, as is the case with Coulomb friction). As $z\GR(\psi, \chi, \nu\1, \nu\2)$ is a central quantity involved in the analytic approximation in Eq.~\eqref{eq:lam}, we plot $z\GR(\psi, \chi\=1, \nu\1\=\nu\2\=0.25)$ in the inset Fig.~\ref{fig:PC_spec} (left) and $z\GR(\psi\=1, \chi, \nu\1\=\nu\2\=0.25)$ in the inset Fig.~\ref{fig:PC_spec} (right). $z\GR$ exhibits a relatively weak dependence on the bimaterial contrast (quantified by either $\chi$ or $\psi$), with values close to unity, though for the same range of $z$ values, the functions $G_{1,2}$ in Fig.~\ref{fig:g_func} exhibit non-negligible variations.

\begin{figure*}[ht]
  \centering
  \includegraphics[width=0.99\textwidth]{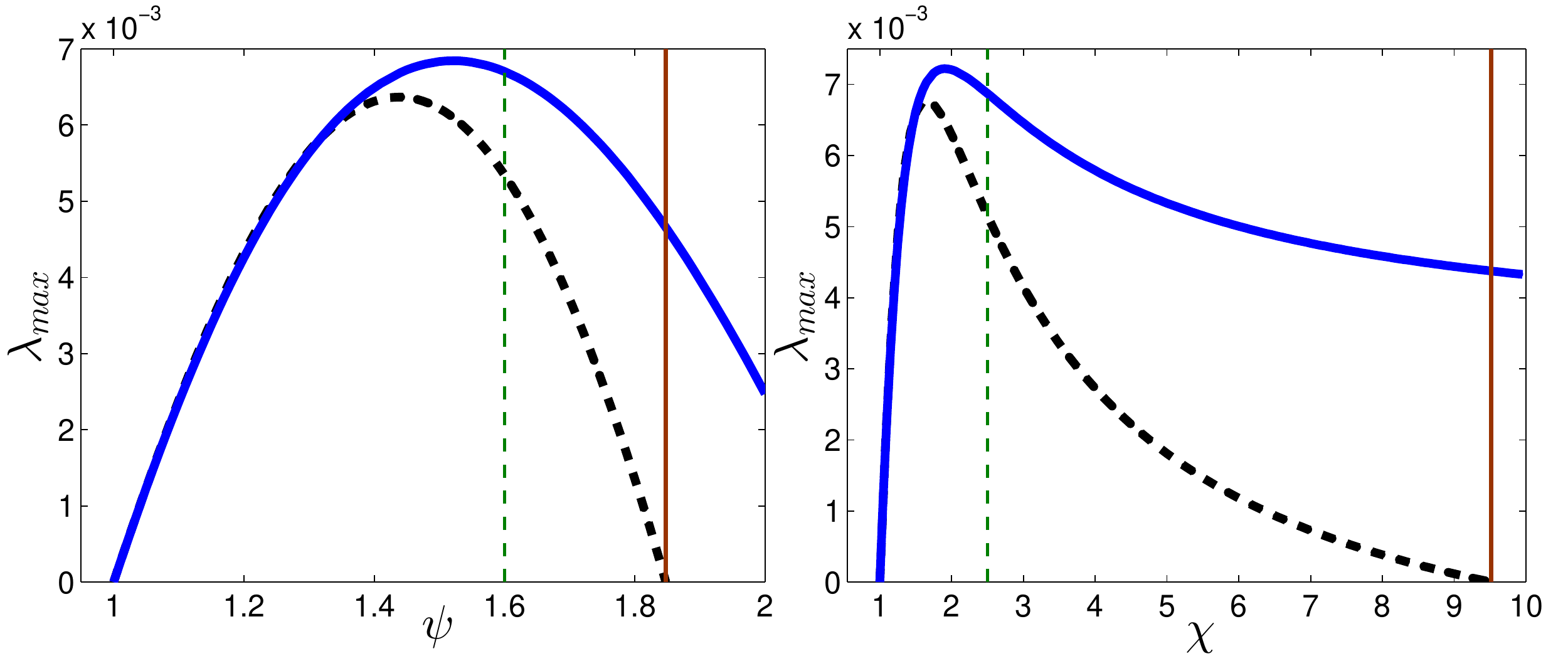}
  \caption{The maximal growth rate $\lambda_{max}$ as a function of $\psi$ for $\chi\!=\!1$ (left, solid line) and as a function of $\chi$ for $\psi\!=\!1$ (right, solid line) obtained from numerical solutions of Eq.~\eqref{eq:spec}. The dashed lines correspond to the analytic approximation in Eq.~\eqref{eq:max}. The vertical dashed lines mark the contrast values considered in Fig.~\ref{fig:PC_spec}. The vertical solid lines mark the maximal contrast for which Generalized Rayleigh waves exist. We used $\nu\1\!=\!\nu\2\!=\!0.25$ and $f\!=\!0.3$ to obtain the results shown in this figure.}
  \label{fig:PC_max_lam}
\end{figure*}

The approximated analytic solution in Eq.~\eqref{eq:lam} allows us to gain analytic insight into the main question posed above, i.e.~whether physical quantities in the considered problem depend monotonically on the bimaterial contrast. The quantity we choose to focus on is the maximal growth rate $\lambda_{max}$ --- a central physical quantity in any linear stability analysis --- which is expected to vanish in the absence of bimaterial contrast (as there is no destabilizing process in the problem in this limit). Using Eq.~\eqref{eq:lam}, we immediately obtain the following estimates for the most unstable mode and the maximal instability growth rate
\begin{equation}
\label{eq:max}
q_{max}\simeq\frac{1}{z\GR}\ ,\qquad\qquad\lambda_{max}\simeq\frac{f G_2(z\GR)}{2\,z\GR\left|G_1'(z\GR)\right|}\ ,
\end{equation}
which are valid in the range of bimaterial contrasts for which $z\GR$ exists (note, though, that while the existence of $z\GR$ is important for the analytic approximation, it does not play a crucial role in the main effect of interest here).
The prediction in Eq.~\eqref{eq:max} shows that indeed $\lambda_{max}$ is proportional to $G_2$ and hence that it vanishes in the absence of a bimaterial contrast, $G_2\=0$, as expected. The dependence of $\lambda_{max}$ on the bimaterial contrast, however, is {\em not exclusively} determined by $G_2$, but rather by the ratio of $G_2(z\GR)$ and $z\GR\!\left|G_1'(z\GR)\right|$, where $G_1$ and $G_2$ depend on the bimaterial contrast (quantified by $\psi$, $\chi$, $\nu\1$, $\nu\2$) both explicitly and implicitly through $z\GR(\psi, \chi, \nu\1, \nu\2)$. The functions $G_2$, $G_1$ and $z\GR$ exhibit a non-trivial dependence on the bimaterial contrast, as shown in Figs.~\ref{fig:g_func} and \ref{fig:PC_spec}, and consequently $\lambda_{max}(\psi, \chi, \nu\1, \nu\2)$ might be non-monotonic despite the fact that $G_2$ is monotonic. Indeed, plotting $\lambda_{max}(\psi, \chi\=1, \nu\1\=\nu\2\=0.25)$ in Fig.~\ref{fig:PC_max_lam} (left) and $\lambda_{max}(\psi\=1, \chi, \nu\1\=\nu\2\=0.25)$ in Fig.~\ref{fig:PC_max_lam} (right), we observe that both quantities exhibit a non-monotonic dependence on the bimaterial contrast. This result explicitly demonstrates that important physical quantities may exhibit a non-monotonic dependence on the bimaterial contrast, i.e.~that there exist some non-trivial contrast conditions in which the frictional bimaterial effect manifests itself most strongly. Note also that the location of the maximal bimaterial effect does not coincide with the disappearance of Generalized Rayleigh waves.

\section{Non-monotonicity of the bimaterial effect in the stability of homogeneous sliding with velocity-strengthening rate-and-state friction}
\label{sec:RSF}

In the previous section we demonstrated that despite the fact that the elastodynamic bimaterial coupling between interfacial slip and normal stress perturbations is a monotonically increasing function of the bimaterial contrast, important physical quantities such as the maximal growth rate of homogeneous sliding instability modes can exhibit a non-monotonic dependence on the bimaterial contrast. This was done for a regularized Coulomb friction constitutive relation. Does this non-monotonicity persist for other, more realistic, interfacial constitutive relations? To address this question we consider in this section sliding along interfaces described by a generic rate-and-state friction constitutive framework. In particular, we consider the following constitutive framework~\citep{Rice1983,Marone1998,Baumberger2006}
\begin{equation}
\tau=f(v,\phi)\,\sigma\qquad\qquad\hbox{and}\qquad\qquad\dot\phi=g\!\left(\!\frac{v\phi}{D}\!\right)\ ,
\label{eq:RSF}
\end{equation}
with $g(1)\=0$ and $g'(1)\!<\!0$. This constitutive framework makes reference to the multi-contact nature of macroscopic frictional interfaces, where $\phi(x,t)$ is an internal state field that quantifies the average lifetime of contact asperities in mesoscopic portions of the interface and $D$ is a memory length related to the linear size of contact asperities. For homogeneous steady state sliding at a slip velocity $V$, we have $\phi\=D/V$ (corresponding to $g(1)\=0$) and $\tau_0\=f(V,D/V)\sigma_0$, where $\sigma_0$ is a constant compressive stress. Since sliding reduces the lifetime of micro-contacts, we have $g'(1)\!<\!0$.

We are interested in the stability of this homogeneous sliding steady state. To that aim, we perturb Eq.~\eqref{eq:RSF} in the form
\begin{equation}
\label{eq:RSF_perb}
\delta\tau = \delta\sigma_{yx}=-f\delta\sigma_{yy}+\sigma_0\delta{f} \ .
\end{equation}
We then substitute the perturbation expressions for $\delta\sigma_{yx}$ and $\delta\sigma_{yy}$ in terms of $\delta\epsilon$ given in Eq.~\eqref{eq:pert}. $\delta{f}/\delta{v}$ is expressed in terms of $\delta{f}/\delta{\phi}$ by taking the variation of $f(v,\phi)$ around steady state~\citep{Ruina1983,Brener2016}. $\delta{f}/\delta{\phi}$ is obtained by taking the variation of $\dot\phi$ in Eq.~\eqref{eq:RSF} around steady state. Finally, using $\delta{v}\=-i\,k\,z\,c_s\1\,\delta\epsilon$, $\delta{f}$ is expressed in terms of $\delta\epsilon$ and once substituted in Eq.~\eqref{eq:RSF_perb} we obtain
\begin{equation}
\gamma \left[G_1(z)-i f G_2(z)\right]+i z\frac{\Delta -i q z}{1-i q z}=0\ ,
\label{eq:specRSF}
\end{equation}
where the dimensionless wavenumber $q\=\frac{k D c_s\1}{|g'(1)|V}$ is {\em different} from the corresponding quantity in Eq.~\eqref{eq:spec} as different normalization is used.

While Eq.~\eqref{eq:spec} depends on a single interfacial parameter $f$, Eq.~\eqref{eq:specRSF} depends on the following three dimensionless interfacial parameters
\begin{equation}
f(v,\phi),\quad \gamma\!\equiv\!\frac{\displaystyle\mu}{\displaystyle\sigma _0\,c_s\1\,\tfrac{\partial{f(v,\phi)}}{\partial{v}}}, \quad \Delta\!\equiv\! \frac{d{f(v,\phi(v))}}{dv}\!\bigg/\frac{\partial{f(v,\phi)}}{\partial{v}} \ ,
\end{equation}
all evaluated at steady state, i.e.~at $v\=V$ and $\phi\=D/V$. $f(v,\phi)$ is simply the friction coefficient at steady state. $\gamma$ quantifies the ratio between the radiation damping coefficient $\mu/c_s\1$~\citep{Rice1993,Rice2001,Crupi2013}, the normal stress $\sigma_0$ and the instantaneous response of the interface to slip velocity variations $\tfrac{\partial{f}}{\partial{v}}$~\citep{Brener2016}. The latter is generically positive, $\tfrac{\partial{f}}{\partial{v}}\!>\!0$, which is also known as the ``direct effect''~\citep{Baumberger2006}. $\Delta$ is the ratio between the steady state variation of the friction coefficient, $\tfrac{df}{dv}$, and $\tfrac{\partial{f}}{\partial{v}}$~\citep{Brener2016}. $\tfrac{df}{dv}$ can be both negative and positive, also for the very same interface at different steady state sliding velocities $V$~\citep{Bar-Sinai2014}. As we would like friction to be stabilizing such that the bimaterial effect is the only destabilizing process in the problem, we choose $\tfrac{df}{dv}\!>\!0$. That is, we perform the analysis for steady state velocity-strengthening friction.

\begin{figure*}[ht]
  \centering
  \includegraphics[width=0.99\textwidth]{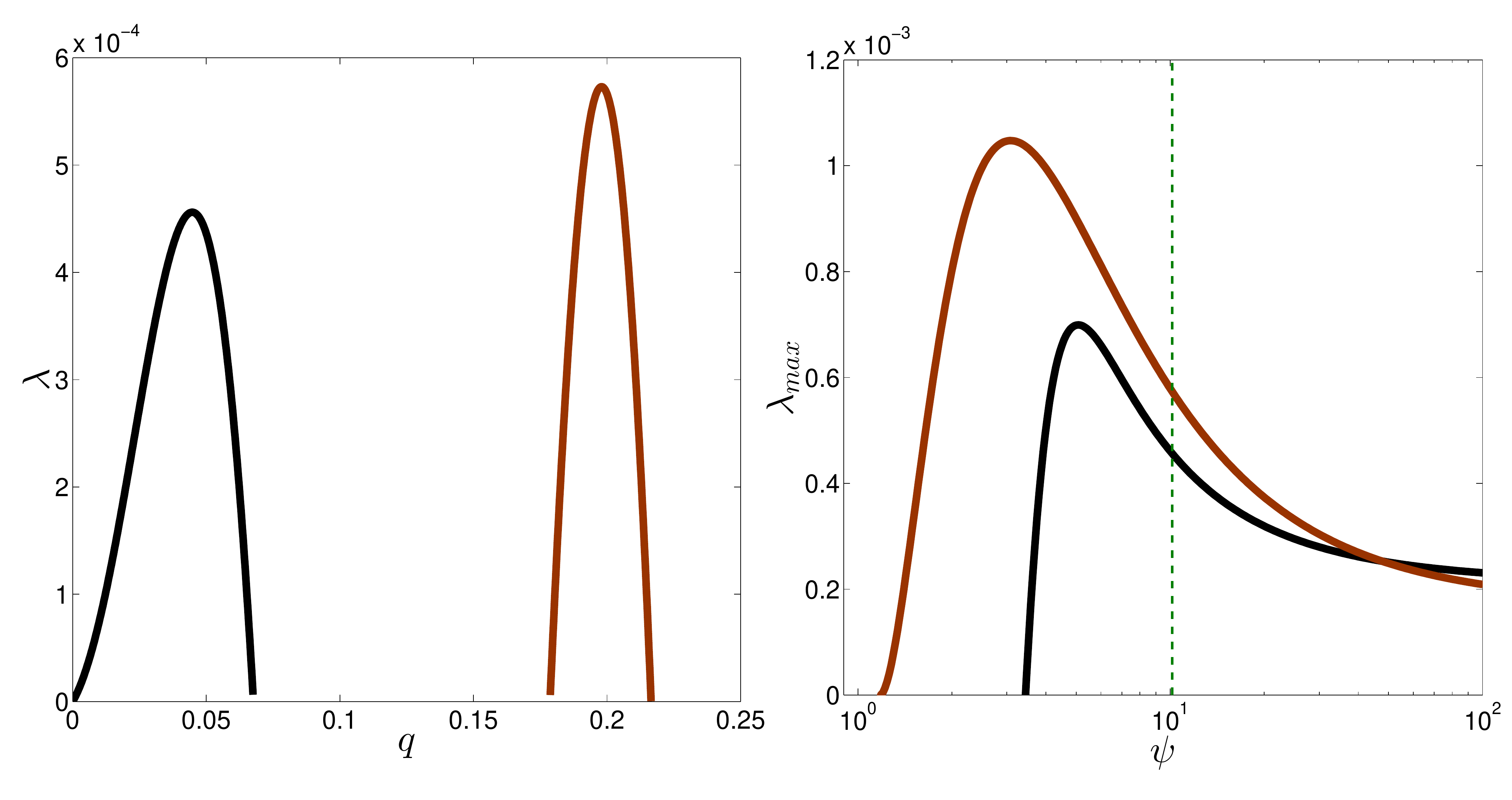}
  \caption{(left) The instability growth rate $\lambda$ as a function of the dimensionless wavenumber $q$ obtained from numerical solutions of Eq.~\eqref{eq:specRSF} with $\psi\!=\!10$, $\chi\!=\!1$, $\gamma\!=\!0.25$, $f\!=\!0.3$, $\Delta\!=\!0.02$, $\nu\1\!=\!0.28$ and $\nu\2\!=\!0.37$. Members of two coexisting families of solutions are shown (capturing different ranges of wavenumbers $q$). (right) The maximal growth rate $\lambda_{max}$ for the two families of solutions of Eq.~\eqref{eq:specRSF} as a function of $\psi$ (the rest of the parameters are as in the left panel). The vertical dashed line corresponds to the instability spectra shown in the left panel ($\psi\!=\!10$). Note that the two unstable ($\lambda_{max}\!>\!0$) families of solutions emerge at different threshold $\psi\!>\!1$ values.}\label{fig:RSF_non_mon}
\end{figure*}

We are now interested in studying the dependence of the solutions of Eq.~\eqref{eq:specRSF} on the bimaterial contrast, where the latter is quantified here by setting $\chi\=1$ and $\nu\1\=\nu\2\=0.25$, and varying $\psi$. Obtaining analytic solutions to Eq.~\eqref{eq:specRSF}, even approximate, is a very serious mathematical challenge and consequently we resort to numerical solutions. In the $\psi\!\to\!1$ limit, i.e.~in the absence of a bimaterial contrast, homogeneous sliding is unconditionally stable because friction is intrinsically stabilizing in this problem. That is, all wavenumbers $q$ feature a negative growth rate $\lambda(q)\=q\Im\left[z\right]\!<\!0$ in the $\psi\!\to\!1$ limit. As $\psi$ is increased, the destabilizing bimaterial effect competes with the stabilizing friction, and beyond a certain level of bimaterial contrast unstable solutions emerge. That is, above a certain threshold value of $\psi$, there exists a range of wavenumbers $q$ for which $\lambda\!>\!0$. Examples of such solutions are shown in Fig.~\ref{fig:RSF_non_mon} (left). In fact, we have found two families (branches) of unstable solutions to Eq.~\eqref{eq:specRSF}, which emerge at different threshold values of $\psi$. Each member of these two families of unstable solutions, as demonstrated in Fig.~\ref{fig:RSF_non_mon} (left), features a maximal growth rate, $\lambda_{max}$.

As explained above, for sufficiently small $\psi$, all perturbation modes are stable. At a threshold value of $\psi$, unstable solutions emerge, where $\lambda_{max}\=0$. As $\psi$ is further increased, $\lambda_{max}$ becomes finite. Is $\lambda_{max}$ a monotonically increasing function of $\psi$? To address this question, we plot in Fig.~\ref{fig:RSF_non_mon} (right) $\lambda_{max}$ as a function of $\psi$, for fixed values of $f$, $\gamma$ and $\Delta$ (see figure caption).
We observe that $\lambda_{max}$ exhibits a non-monotonic dependence on the bimaterial contrast $\psi$, similarly to the non-monotonic behavior observed in the regularized Coulomb friction analysis presented in Fig.~\ref{fig:PC_max_lam}. Therefore, for two very different interfacial constitutive relations --- a regularized Coulomb friction and velocity-strengthening rate-and-state friction --- important physical quantities exhibit a non-monotonic dependence on the bimaterial contrast. While we cannot rigorously prove that any physical quantity of interest for a general interfacial constitutive relation exhibits a non-monotonic dependence on the bimaterial contrast, we believe that the results presented in sections \ref{sec:Coulomb} and \ref{sec:RSF} provide strong evidence that this might be the case quite generically. Next, we explore the same question in the context of unsteady nonlinear rupture propagation.

\section{Non-monotonicity of the bimaterial effect in dynamic rupture simulations with slip-weakening friction}
\label{sec:dynamics}

The analyses presented in the previous sections focussed on the linear stability of steady state homogeneous sliding along bimaterial interfaces. Our goal here is to extend these analyses to nonlinear rupture propagation, which is relevant to a broad range of frictional and earthquake problems. To that aim, we performed simulations of in-plane dynamic rupture between two half-spaces using the 2D spectral element code SEM2DPACK~\citep{Ampuero2002}. As above, the upper half-space features elastic moduli $\mu\1$ and $\nu\1$, and lower half-space has $\mu\2$ and $\nu\2$. The upper half-space is taken to be more compliant, $\mu\1\!<\!\mu\2$, and we focus on rupture propagating in the slip direction of the more compliant half-space, referred to as the positive direction.

The interface between the two half-spaces is characterized by a linear slip-weakening friction law, which is a minimal constitutive framework to account for the transition from stick to slip involved in rupture propagation. In particular, we have
\begin{equation}
\label{eq:friction law}
\tau \le f(\epsilon)\,\sigma^* \ ,
\end{equation}
where the inequality holds under stick conditions ($\epsilon\=0$) and the equality holds under slip conditions ($\epsilon\!>\!0$). $f(\epsilon)$ is given by
\begin{equation}
 f(\epsilon)=\begin{cases}
                 f_s -(f_s-f_d)\epsilon/D_{\hbox{\tiny{SW}}} & 0\le \epsilon \le D_{\hbox{\tiny{SW}}}\\
                 f_d & \epsilon>D_{\hbox{\tiny{SW}}}\\
                 \end{cases}\ ,
\label{eq:SW}
\end{equation}
where $f_s$ and $f_d$ are the static and dynamic friction coefficients, respectively, and $D_{\hbox{\tiny{SW}}}$ is a characteristic slip-weakening distance that controls the transition from $f_s$ to $f_d$. The regularized normal stress $\sigma^*$ satisfies
\begin{equation}
\label{eq:regularization}
\dot\sigma^* = -\frac{v + v^*}{L} (\sigma^* - \sigma) \ ,
\end{equation}
which, similarly to Eq.~\eqref{eq:PC}, ensures that the problem is well-posed. $v^*$ is chosen to be much smaller than the characteristic slip velocity near the rupture front and the regularization length $L$ satisfies $L\!\ll\!D_{\hbox{\tiny{SW}}}$ such that $\sigma^*$ evolves on a timescale shorter than that of the slip-weakening process, making the latter largely independent of the normal stress regularization in Eq.~\eqref{eq:regularization}. In the slipping region, for $\epsilon\!>\!D_{\hbox{\tiny{SW}}}$, the model bears close similarity to the regularized Coulomb friction model analyzed under steady state sliding conditions in section~\ref{sec:Coulomb}.

The bimaterial contrast is controlled in our simulations by varying the shear wave-speeds ratio $\zeta\=c_s\2/c_s\1\!>\!1$. This is realized in the following manner: for a given $\zeta$, we first set $\chi\=\zeta^{-1}$. Since $\zeta\=\sqrt{\psi\chi}$, we have $\psi\=\chi^{-3}$. Finally, we set $\nu\1\=\nu\2\=0.25$, use $\mu\2\=\psi\mu\1$ and select $\mu\1$ such that $\bar{\mu}_0/\mu\0\!\equiv\!2\mu\1 G_1(z\!\to\!0)/\mu\0$ attains a fixed value of $4/3$ (the value of $(1-\nu)^{-1}$ for $\nu\=0.25$). Here, $\mu\0$ is a reference shear modulus in the absence of bimaterial contrast, which is used to normalize quantities of stress dimensions, and $\bar{\mu}_0(\mu\1, \nu\1, \mu\2, \nu\2)$ represents the effective shear modulus for the bimaterial problem~\citep{Rubin2007}. The motivation behind keeping $\bar{\mu}_0$ constant, while varying $\zeta$, $\chi$ and $\psi$ as just explained, is that it may help reducing trivial dependencies on the bimaterial contrast and hence revealing potentially less trivial dependencies. In particular, it has been suggested that $\bar{\mu}_0$ affects the rupture nucleation length $L_c$ in the bimaterial problem in much the same way as the ordinary shear modulus $\mu$ affects it in the homogeneous materials problem~\citep{Rubin2007}. Consequently, rupture in our simulations is initiated by over-stressing a nucleation zone of length $L_c$ slightly above the static friction level. The nucleation length $L_c$ is estimated using the Uenishi-Rice length for homogeneous material interfaces~\citep{Uenishi2003}, where the combination $\mu/(1\!-\!\nu)$ of the ordinary shear modulus $\mu$ and Poisson's ratio $\nu$ is replaced by $\bar{\mu}_0$~\citep{Rubin2007}. That is, we use $L_c\=\frac{1.158\,\bar{\mu}_0 D_{\hbox{\tiny{SW}}}}{\sigma_0 (f_s-f_d)}$, where $\sigma_0$ is the magnitude of the applied normal stress, as above. Outside the nucleation zone, an initial shear stress of magnitude $\tau_0$, which lies between the static and dynamic friction levels, is applied.

In our simulations, velocities and stresses are measured relative to the shear wave-speed and the reference shear modulus $\mu\0$ in the absence of bimaterial contrast ($\zeta\=1$), respectively, and lengths are measured relative to the slip-weakening length $D_{\hbox{\tiny{SW}}}$. In these units, we set $L\=0.2$ and $v^*\=0.01$, in addition to $f_s\=0.6$ and $f_d\=0.3$. The initial shear stress in the nucleation zone is set to $0.61\sigma_0$ and the shear stress outside the nucleation zone to $\tau_0\=0.4071\sigma_0$. Finally, the nucleation length is set to $L_c\=5.16$ and $x\=0$ is located at the middle of the nucleation zone. We follow rupture propagation in the positive direction by tracking the spatiotemporal evolution of both the slip velocity $v(x,t)$ and the magnitude of the normal stress reduction $\Delta\sigma(x,t)\!>\!0$, where $t\=0$ is the nucleation time, for various bimaterial contrasts $\zeta\!\ge\!1$.
\begin{figure*}[ht]
\centering
  \includegraphics[width=1.04\textwidth]{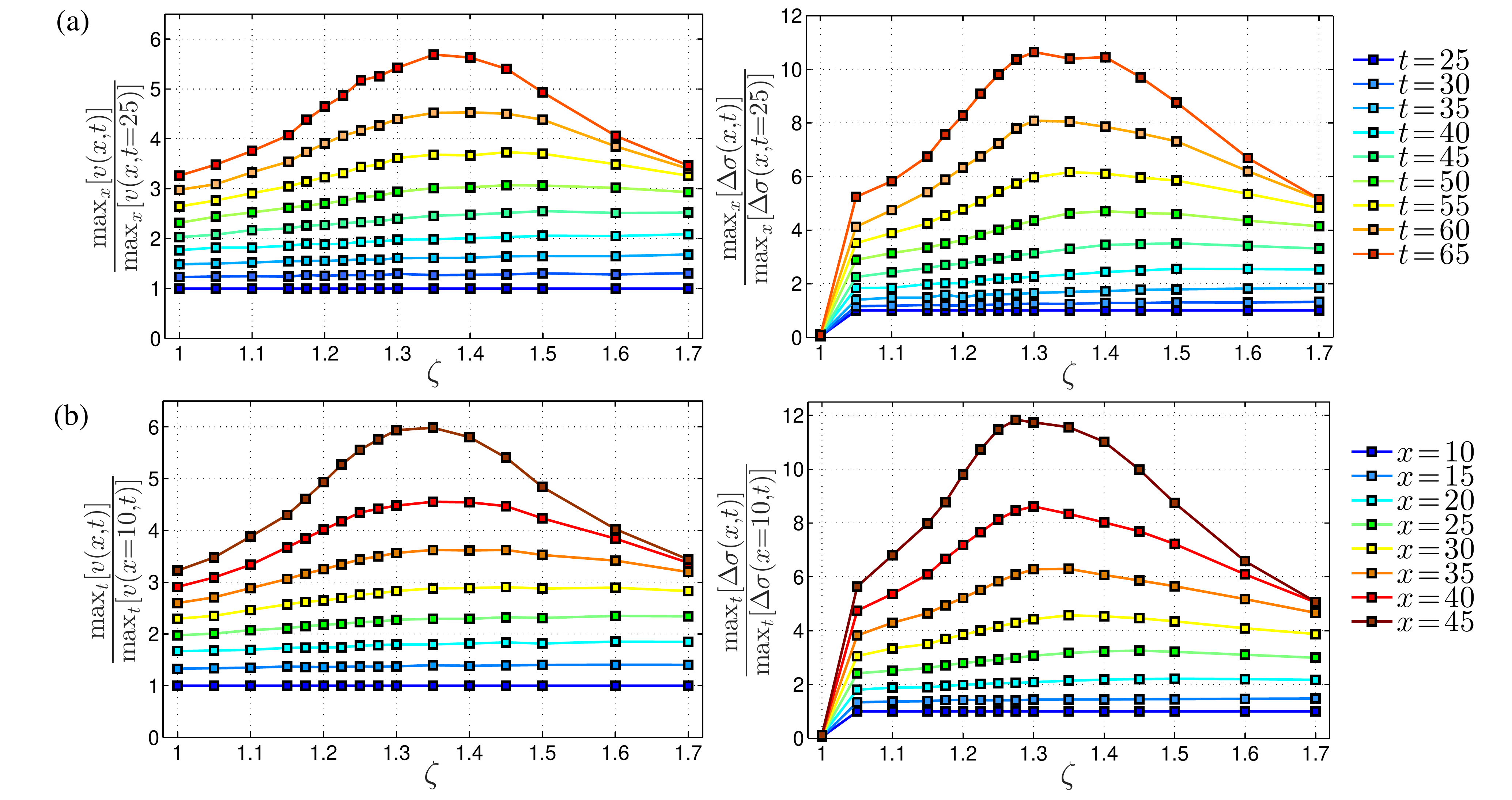}
  \caption{Results of in-plane rupture simulations using linear slip-weakening friction (see text for details). (a) The spatial maximum of the slip velocity $v(x,t)$ at different times (see legend on the right), $\max_x[v(x,t)]$, normalized by $\max_x[v(x,t\!=\!25)]$, as a function of $\zeta$ (left). The spatial maximum of the normal stress reduction $\Delta\sigma(x,t)$ at different times (see legend on the right), $\max_x[\Delta\sigma(x,t)]$, normalized by $\max_x[\Delta\sigma(x,t\!=\!25)]$ (except for $\zeta\!=\!1$, where $\Delta\sigma\!=\!0$ due to the absence of a bimaterial contrast), as a function of $\zeta$ (right). (b) The temporal maximum of the slip velocity (left) at different fixed locations on the fault (corresponding to a fixed rupture propagation distance in the positive direction, see legend on the right), $\max_t[v(x,t)]$, and of the normal stress reduction (right), $\max_t[\Delta\sigma(x,t)]$, both normalized by the corresponding values at $x\!=\!10$, as a function of $\zeta$.}
\label{fig:rupture}
\end{figure*}

Our goal now is to define rupture-related quantities that are sensitive to the bimaterial effect in order to study their dependence on the bimaterial contrast. Unlike the analyses presented in previous sections, where the dependence on the bimaterial contrast of various time-independent quantities --- such as the instability growth rate --- was cleanly defined, rupture propagation is an intrinsically time-dependent process that in many cases does not reach steady state. Hence, the bimaterial contrast may affect different stages of rupture, such as nucleation and post-nucleation propagation style (e.g.~sub-shear vs.~super-shear, fault opening vs.~no fault opening), where early-time dynamics affect later times. To minimize these potentially intervening effects, we (i) use a nucleation length that corresponds to a fixed effective shear modulus $\bar{\mu}_0$ (as explained above) (ii) limit ourselves to the regime where no fault opening takes place and (iii) force the physical quantities of interest to be independent of $\zeta$ at some point in time or at a fixed rupture propagation distance by a proper normalization choice.

In particular, we calculate the spatial maximum of the slip velocity $v(x,t)$ at different times, $\max_x[v(x,t)]$, and normalize it by $\max_x[v(x,t\=25)]$. Therefore, by construction, this normalized quantity equals unity independently of $\zeta$ at $t\=25$. The $\zeta$ dependence of this normalized quantity as time progresses, $t\!>\!25$, is plotted in Fig.~\ref{fig:rupture}a (left). It is observed that $\max_x[v(x,t)]/\!\max_x[v(x,t\=25)]$ becomes a non-monotonic function of $\zeta$ with increasing time. The maximum of this function, which exhibits a rather weak dependence on time once it appears, occurs at a $\zeta$ value which is close to the prediction in Fig.~\ref{fig:PC_max_lam} (recall that $\zeta\=\sqrt{\psi\chi}$ and that $\chi\=1$ was used there). In Fig.~\ref{fig:rupture}a (right), we plot the spatial maximum of the normal stress reduction $\Delta\sigma(x,t)$ at different times, $\max_x[\Delta\sigma(x,t)]$, normalized by $\max_x[\Delta\sigma(x,t\=25)]$ (except for $\zeta\=1$, where $\Delta\sigma\=0$ due to the absence of a bimaterial contrast), as a function of $\zeta$. It is observed that $\max_x[\Delta\sigma(x,t)]/\!\max_x[\Delta\sigma(x,t\=25)]$ becomes a non-monotonic function of $\zeta$ with increasing time, attaining a maximum at a contrast level similar to the one observed in Fig.~\ref{fig:rupture}a (left). In Fig.~\ref{fig:rupture}b, we plot the temporal maximum of the slip velocity (left) at different fixed locations on the fault (corresponding to a fixed rupture propagation distance in the positive direction), $\max_t[v(x,t)]$, and of the normal stress reduction (right), $\max_t[\Delta\sigma(x,t)]$, both normalized by the corresponding values at $x\=10$, as a function of $\zeta$. Both quantities develop a non-monotonic dependence on $\zeta$ with increasing rupture propagation distance. All in all, the results presented in Fig.~\ref{fig:rupture} provide evidence that the non-monotonicity of the bimaterial effect persists in non-steady, transient frictional dynamics.

The growth rate of both slip velocity and changes in normal stress in Fig.~\ref{fig:rupture} are peaked at $\zeta$ values between $1.3$ and $1.4$, rather than at the maximal used value of $1.7$. This is similar to the results of~\citet{Ben-Zion1998} on the amplification of slip velocity with propagation distance vs.~bimaterial contrast in simulations with a constant Coulomb friction. The non-monotonicity observed in Fig.~\ref{fig:rupture} can be at least partially rationalized through an analytic estimation developed in~\citet{Rubin2007} for linear slip-weakening friction. There, cf.~Eq.~(12), the normal stress reduction is expressed as a function of $\bar{\mu}/\mu^*\=2G_1(z\=-v_r/c_s\1)/G_2(z\=-v_r/c_s\1)\!\equiv\!{\cal G}$ in the approximated form $\Delta\sigma\!\sim\!({\cal G}-f)^{-1}$, where $v_r$ is the rupture propagation velocity. For small rupture velocities, ${\cal G}$ is positive and larger than $f$, and is a decreasing function of the bimaterial contrast. This implies that $\Delta\sigma$ is a monotonically increasing function of the contrast for small rupture velocities. For large rupture velocities, relevant for the rupture propagation velocities $v_r$ in the simulations analyzed in Fig.~\ref{fig:rupture}, ${\cal G}$ can equal $f$ --- say at $\zeta_f$ --- where the expression for $\Delta\sigma$ formally diverges. For $\zeta\!>\!\zeta_f$, ${\cal G}\!>\!f$ and is an increasing function of the bimaterial contrast. This implies that $\Delta\sigma$ for large velocities is a monotonically decreasing function of the contrast for $\zeta\!>\!\zeta_f$, consistent with the decreasing functions of $\zeta$ observed in Fig.~\ref{fig:rupture} after the peak value. For $\zeta\!<\!\zeta_f$, ${\cal G}$ is temporarily larger than $f$ due to early-time slow rupture acceleration, but may approach $f$ at later stages of rupture evolution. This is consistent with the delayed growth of $\Delta\sigma$ for small $\zeta$ observed in Fig.~\ref{fig:rupture} to the left of the peak value. Taken together, the analytic estimate $\Delta\sigma\!\sim\!({\cal G}-f)^{-1}$ may rationalize the non-monotonicity observed in Fig.~\ref{fig:rupture}.

\section{Summary and discussion}
\label{sec:summary}

The analyses performed in this paper provide a series of analytic and numerical results on the dependence of various quantities related to the dynamics of bimaterial frictional interfaces on the degree of contrast across the interface. The results demonstrate that while the bimaterial coupling between interfacial slip and normal stress perturbations is a monotonically increasing function of the bimaterial contrast, when an interfacial constitutive relation (friction law) is considered, various physical quantities exhibit a non-monotonic dependence. This seemingly generic effect emerges both in the context of the stability of steady state sliding and unsteady rupture propagation, using several interfacial constitutive laws including regularized Coulomb friction, rate-and-state friction and slip-weakening friction.

The existence of a non-monotonic dependence of various physical quantities on the bimaterial contrast implies that there exists a level of bimaterial contrast for which the destabilizing bimaterial effect is maximal. The exact contrast level for maximal bimaterial effect may vary from one physical quantity to another and may depend on the interfacial constitutive relation. The existence of a maximum may have significant implications for interfacial dynamics and rupture, since the magnitude of the bimaterial effect strongly affects the degree of dynamic changes in normal stress and elastodynamic frictional weakening. This, in turn, controls the stability of sliding, style of rupture, frequency range and spatial pattern of radiated waves, amount of frictional heat dissipation accompanying interfacial rupture, and other issues discussed in~\citet{Ben-Zion1998} and later papers.

The contrast of seismic velocities across the seismogenic sections (e.g.~depth $>\!3$ km) of large natural faults is typically $10-20\%$ or less~\citep[e.g.][]{Ben-Zion1992,Lewis2007,Allam2014}. The velocity contrasts between the edges of core damage zones of faults and the surrounding rock may be $50\%$ or more, but these damage zones are usually limited to the top few km of the crust~\citep[e.g.][] {Ben-Zion2003, Yang2011}. In the context of laboratory experiments, where the bimaterial contrast is externally controlled by the selection of the pair of materials to be used, our results can guide the selection of a bimaterial contrast so as to maximize the sensitivity to the bimaterial effect. Laboratory experiments can also systematically test our predictions by controllably varying the bimaterial contrast. It is important to note in this context that the destabilizing bimaterial effect and related dynamic phenomena evolve with propagation distance and time~\citep[e.g.][]{Adams1995,Andrews1997,Ben-Zion2001,Ranjith2001}. Focusing on the rupture propagation problem, prominent bimaterial effects as observed in Fig.~\ref{fig:rupture} require either large rupture velocity or long propagation distance~\citep{Ben-Zion2002}, which in general depend on additional factors such as rupture nucleation, background stress and strain rate, and fault constitutive laws~\citep[e.g.][]{Shi2006,Ampuero2008a,Scala2017}. In the context of glacial earthquakes with a large property contrast between ice and rock, our results may partly explain why the degree of slip instability is often limited compared to typical tectonic earthquakes~\citep{Wiens2008}, though other factors (e.g.~loading style, ice melting, drainage) can also play an important role.

Some of the analyses performed in this study can be extended to other configurations producing dynamic changes in normal stress along a frictional interface. Examples include contrast in poroelastic properties across faults~\citep{Dunham2008} and asymmetric geometrical properties of the solids across faults~\citep{Aldam2016}, where the latter are directly relevant to geometric variations and free surface effects on dipping faults~\citep{Oglesby1998, Gabuchian2017}. In this context, it is interesting to note that several of these effects, and the elastodynamic bimaterial effect analyzed in this study, can coexist at the shallow portions of subduction zones. Moreover, the degree of geometric and/or bimaterial contrast across the subduction plate interface can vary as a function of distance towards the trench. These factors may complicate the interplay between dynamic change of normal stress and friction along the plate interface, and the displacement partitioning between the overriding plate and the subducting plate.

Finally, we emphasize that our results on the non-monotonic dependence of various quantities on the bimaterial contrast are based on separate analyses of the stability of steady sliding against linear perturbations and of non-steady nonlinear rupture propagation, focusing primarily on the sub-shear regime. Future theoretical and experimental studies can improve the understanding of the dynamics leading to non-monotonicity and other aspects of bimaterial ruptures by considering additional ingredients known to affect rupture behavior, such as nucleation process and properties of the initial stress field, and examining results also for the super-shear rupture regime and negative propagation direction.

\section*{Appendix: The elastodynamic transfer functions $G_1$ and $G_2$}
\label{app:G_func}

The elastodynamic transfer functions $G_1$ and $G_2$, introduced in section~\ref{sec:bimaterial} and used throughout the paper, take the following form
\begin{widetext}
\begin{eqnarray}
\label{eq:G_func}
G_1\=\frac{\chi  \psi ^2 z^2 R\1 R\2 \left(\chi  \psi ^2 R\2 \alpha\1_d+R\1 \alpha\2_d\right)}{z^4 \left(\chi  \psi ^2 R\2 \alpha\1_d+R\1 \alpha\2_d\right) \left(\chi  \psi ^2 R\2 \alpha\1_s+R\1 \alpha\2_s\right)-\chi ^2 \psi ^2 \left(P\2 R\1-\psi  P\1 R\2\right)^2}\ ,\nonumber\\
G_2\=
\frac{\chi ^2 \psi ^3 R\1 R\2 \left(P\2 R\1-\psi  P\1 R\2\right)}{z^4 \left(\chi  \psi ^2 R\2 \alpha\1_d+R\1 \alpha\2_d\right) \left(\chi  \psi ^2 R\2 \alpha\1_s+R\1 \alpha\2_s\right)-\chi ^2 \psi ^2 \left(P\2 R\1-\psi  P\1 R\2\right)^2}\ ,
\nonumber
\end{eqnarray}
\end{widetext}
where $P\n\!\equiv\!2\alpha\n_d\alpha\n_s-\left({\alpha\n_s}^2+1\right)$ and $R\n\!\equiv\!4\alpha\n_d\alpha\n_s-\left({\alpha\n_s}^2+1\right)^2$, with $n\=1,2$.

\acknowledgments

E.~B.~acknowledges support from the Israel Science
Foundation (Grant No.~295/16), the William Z. and Eda
Bess Novick Young Scientist Fund, COST Action MP1303,
and the Harold Perlman Family. M.~A.~acknowledges Yohai Bar-Sinai for helpful guidance and assistance. We thank Hadar Shlomai for pointing out useful references. We acknowledge the use of the 2D spectral element code SEM2DPACK available at: \url{https://sourceforge.net/projects/sem2d/}.


\end{document}